**TITLE**:

An AI tool for automated analysis of large-scale unstructured clinical cine CMR databases

**BRIEF TITLE**:

AI tool for analysis of clinical cine CMR databases

**AUTHORS:**

Jorge Mariscal-Haran, *PhD*,[a], Clint Asher, *MD*,[a,b], Vittoria Vergani, *MD*,[a], Maleeha Rizvi, *MD*,[a,b], Louise Keehn, *PhD*,[c], Raymond J. Kim, *MD*,[d], Robert M. Judd, *PhD*,[d], Steffen E. Petersen, *MD*, *PhD*,[e,f,g,h], Reza Razavi, *MD*, *PhD*,[a,b], Andrew P. King, *PhD*,[a], Bram Ruijsink, *MD*, *PhD*,[a,b,i,*], Esther Puyol-Antón, *PhD*,[a,*]

* Shared last authors

**AFFILIATIONS:**

a) School of Biomedical Engineering and Imaging Sciences, King's College London, London, UK
b) Department of Adult and Paediatric Cardiology, Guy's and St Thomas' NHS Foundation Trust, London, UK
c) Department of Clinical Pharmacology, King's College London British Heart Foundation Centre, St Thomas' Hospital, London, UK
d) Division of Cardiology, Department of Medicine, Duke University, Durham, North Carolina, USA
e) National Institute for Health Research (NIHR) Barts Biomedical Research Centre, William Harvey Research Institute, Queen Mary University London, London, UK
f) Barts Heart Centre, St Bartholomew's Hospital, Barts Health NHS Trust, London, UK
g) Health Data Research UK, London, UK
h) Alan Turing Institute, London, UK
i) Department of Cardiology, Heart and Lung Division, University Medical Center Utrecht, Utrecht, The Netherlands

**ADDRESS FOR CORRESPONDENCE:**

Esther Puyol-Antón

The Rayne Institute

4th Floor Lambeth Wing, St Thomas Hospital

Westminster Bridge Road

SE1 7EH London, United Kingdom

E-mail: esther.puyol_anton@kcl.ac.uk

**WORD COUNT**: 5,266 words

**NO. OF TABLES:** 3

**NO. OF FIGURES:** 6

**SUPPLEMENTARY METHODS:** 8

**SUPPLEMENTARY TABLES:** 8

**SUPPLEMENTARY FIGURES:** 4

**ABSTRACT:**

**Introduction:** Artificial intelligence (AI) techniques have been proposed for automating analysis of short axis (SAX) cine cardiac magnetic resonance (CMR), but no CMR analysis tool exists to automatically analyse large (unstructured) clinical CMR datasets. We develop and validate a robust AI tool for start-to-end automatic quantification of cardiac function from SAX cine CMR in large clinical databases.

**Methods:** Our pipeline for processing and analysing CMR databases includes automated steps to identify the correct data, robust image pre-processing, an AI algorithm for biventricular segmentation of SAX CMR and estimation of functional biomarkers, and automated post-analysis quality control to detect and correct errors. The segmentation algorithm was trained on 2793 CMR scans from two NHS hospitals and validated on additional cases from this dataset (n=414) and five external datasets (n=6888), including scans of patients with a range of diseases acquired at 12 different centres using CMR scanners from all major vendors.

**Results:** Median absolute errors in cardiac biomarkers were within the range of inter-observer variability: <8.4mL (left ventricle volume), <9.2mL (right ventricle volume), <13.3g (left ventricular mass), and <5.9% (ejection fraction) across all datasets. Stratification of cases according to phenotypes of cardiac disease and scanner vendors showed good performance across all groups.

**Conclusions:** We show that our proposed tool, which combines image pre-processing steps, a domain-generalisable AI algorithm trained on a large-scale multi-domain CMR dataset and quality control steps, allows robust analysis of (clinical or research) databases from multiple centres, vendors, and cardiac diseases. This enables translation of our tool for use in fully-automated processing of large multi-centre databases.

# Introduction

Cardiac Magnetic Resonance (CMR) is the gold standard for biventricular volume and function quantification as well as myocardial tissue characterisation [1]. In recent years, artificial intelligence (AI) methods have achieved human-level accuracy in the segmentation of short-axis cine CMR [2–6]. A major advantage of AI is the automation of CMR analysis, which could unlock large quantities of new data for clinical research and to inform care. Automated AI-based tools now exist for analysis of clinical CMR at the point of acquisition [4, 7] as well as retrospectively from highly structured and controlled databases (e.g. our previously developed AI-CMR$^{QC}$ tool [2]). However, many large clinical and research CMR databases remain underexploited because of the work involved in making the data suitable for use by AI. A number of tasks are crucial to being able to exploit such data but are often overlooked. For example, selecting the required scans from a wider range of acquisitions, processing the data to ensure consistency of data format/structure, checking the data for quality and dealing with inconsistencies in image acquisition and labelling protocol (e.g. presence or absence of myocardial segmentations for end-systolic images, including or excluding papillary muscles). In this paper, we address these challenges and in the process make two major contributions to the field of AI-based CMR analysis.

## Contributions

Our first major contribution is to present the first example of a fully automated AI-based start-to-end pipeline with quality control (QC) for analysis of large-scale unstructured databases of short axis (SAX) CMR. To achieve this, we adapt a state-of-the-art AI segmentation model to enable it to deal with the variations in labelling protocol found in many clinical imaging databases. We embed this model into a robust pipeline featuring automated selection of SAX CMR scans, segmentation and estimation of a range of functional biomarkers followed by automated QC. We train and evaluate the framework on a large and heterogeneous database of CMR scans with ground truth segmentations. This last point is made possible by our second major contribution. We describe the first example of fully automated quality assurance (QA) and correction of *ground truth* segmentations. It is often overlooked that many clinical and research databases feature ground truth segmentations that contain errors. To address this, our QA tools automatically flag cases with potential errors in their ground truth segmentations and even correct some of them automatically.

We perform extensive internal and external validation of our pipeline on datasets featuring a wide range of diseases and scanner types unseen during training, demonstrating that it can automatically process large unstructured databases in a robust way.

# Methods

## Datasets

Our proposed CMR analysis tool was developed and validated using routine clinical CMR scans analysed by expert CMR cardiologists from two NHS hospitals (n=3207), which we refer to as the "NHS" dataset. Every CMR study was annotated manually by clinical fellows (n=12), following the standard operating procedures as defined in the SCMR guidelines [23]. All segmentations were reviewed by level 3 accredited CMR consultants (n=4). The acquisition, pre-processing, and characteristics of this dataset are detailed in Supplemental Method A and Supplemental Figure 1. Additionally, the tool was validated externally on the "Duke" clinical dataset from Duke Cardiovascular Magnetic Resonance Center, Durham, NC (n=1319); and on four public datasets of research CMR scans: UK Biobank or "UKBB" [8] (n=4872), "ACDC" [9] (n=150), "M&Ms" [10] (n=375), and "M&Ms-2" [11] (n=360). For all datasets, short-axis CMR images acquired at end-diastole (ED) and end-systole (ES) were available. Across the datasets, variation existed in segmentation protocol. All public datasets feature manual or semi-automated segmentations at ED and ES of the right ventricle (RV) blood pool, left ventricle (LV) blood pool and LV myocardium. However, in both clinical datasets (NHS and Duke), the RV and LV were not always segmented in the same frame and myocardial segmentations were not always present (particularly in ES). In the Duke dataset, papillary muscles were excluded from the LV and RV blood pools whereas in the other datasets they were included. We note that these types of variation are common in real-world datasets but AI techniques to handle them are currently lacking. The dataset characteristics are summarised in Table 1. Additionally, disease and scanner information were stored for each CMR case – where available – to perform a stratified analysis (see Supplemental Table 1 for additional information on the different scanners and image acquisition protocols).

## Quality assessment

To ensure the quality of the ground truth data used for training and validation, a thorough data quality assessment ($QA^{gt}$) was applied to all manual segmentations. This process was automated and contained a number of common steps with our QC of model outputs that will be described later. Firstly, to allow validation of the data against clinical metrics, including stroke volume (SV) and ejection fraction (EF), exams in which only one frame (ED or ES) was segmented were excluded. Secondly, we developed an automated algorithm based on data- and knowledge-based criteria to flag anatomical abnormalities in the segmentations (See Supplemental Method B). In essence, these criteria check whether the acquisition and/or segmentation covers the heart from base to apex, the size of the segmentations, the existence of parts of the segmentation which are disconnected from the RV and LV blood pools, gaps in segmentations, and discordance between LV and RV segmentation size and stroke volume (SV). Thirdly, duplicate segmentations were flagged. Such duplication may occur when the clinician mistakenly uploads the same ground truth segmentation multiple times, or when an original segmentation is corrected by a reviewer.

Subsequent to the automated $QA^{gt}$, all flagged segmentations underwent a manual review by four expert CMR cardiologists to decide inclusion or exclusion of the data from the training and validation data. The outcome of this manual quality assessment of ground truth segmentations ($QA^{gt}$) is detailed in Supplemental Method C and a breakdown of cases excluded by $QA^{gt}$ is reported in Supplemental Table 2. Note that this manual review of flagged case was only necessary to ensure the high fidelity of our training data, and is not required when processing test data. The pipeline for processing new test data is fully automated.

## Automated CMR analysis tool

Our proposed CMR analysis tool consists of CMR image pre-processing steps, an AI method that automatically selects the cine acquisitions prior to image analysis, an AI method that segments the ventricles and the myocardium from short-axis cine CMR stacks, and a post-analysis QC step (see Figure 1). The tool outputs the segmentations for all frames of the full CMR short axis stack together with cardiac biomarkers such as ventricular volumes and ejection fraction.

### Step 1: CMR pre-processing

The first step automatically converts CMR images from their native format to a common format for subsequent analysis (see Supplemental Method D and Supplemental Table 3). This is an important but often-overlooked step, as a large variation in file format exists (e.g. different DICOM headers), which challenges the accurate translation of critical metadata, including voxel sizes, slice spacings and temporal information.

## Step 2: Data identification

Subsequently, we use our previously published image classification framework [12] to automatically identify the different cine short-axis (SAX) CMR sequences from the data and format them into a unified structure for analysis. This step was only performed for the two clinical databases, NHS and Duke, as the other databases only provided the SAX CMR sequence.

## Step 3: CMR segmentation

The LV and RV endocardium and the LV myocardium are segmented from SAX cine CMR using "nnU-Net", a state-of-the-art medical imaging segmentation framework [13]. nnU-Net aims at reducing the effect of heterogeneities inherent in imaging data (in this study, CMR data) from different clinical centres, MRI vendors, or imaging protocols. To do so, nnU-Net automatically adapts its image pre-processing (z-score intensity normalisation and image resampling), network architecture, and hyperparameters to any given image dataset. These strategies allow nnU-Net to outperform most AI methods (even highly specialised ones) in international medical image segmentation challenges [13].

Additionally, we modified nnU-Net's loss function to tackle the problem of inconsistent image labelling protocols. For example, in most clinical contexts, the myocardium is only segmented at the ED but not in the ES frame. More details on nnU-Net and our loss function adaptation are provided in Supplemental Method E.

A further step was implemented to allow exclusion of papillary muscles from the LV and RV blood pool segmentations using Otsu's threshold method [14] (see Supplemental Method F more details). This step can be applied per clinician preference. Here, it was used for the Duke dataset, to cater for the segmentation protocol followed when forming the ground truth data for the Duke dataset (see Supplemental Figure 2).

## Step 4: Post-analysis QC

AI algorithms typically perform well in pixelwise segmentation of images but lack biophysical/anatomical constraints. This means that segmentation outputs are not always realistic. For example, a cardiac segmentation algorithm could output a result that includes an additional region of blood pool/myocardium outside of the heart (e.g. in the stomach area, which can appear like a muscular wall with fluid inside), or create a hole in the LV myocardial lateral wall. We developed a set of automated QC steps[1] that detect and potentially address these errors in segmentations, using prior knowledge of the anatomy and physiology of the heart. A number of these steps also feature in the $QA^{gt}$ of the databases used for model training (see Supplemental Methods B and G). On the output of the CMR segmentation model, the post-analysis segmentation QC detects segmentations that (i) are discontinuous (e.g., non-segmented slices between segmented slices), (ii) do not adhere to the anatomical relationship between the ventricles and the myocardium, (iii) are disconnected from the rest of the heart. See Supplemental Method G for details. Where possible, segmentations were automatically adapted after being flagged by these QC steps (e.g. segmentations outside the heart area were automatically deleted). More complex issues (for example gross distortions in anatomy or large differences in (stroke) volumes between ventricles) were not addressed. However, those cases were still flagged by the QC step for clinician review (see Supplemental Method G and Supplemental Table 4).

[1] Note the distinction between QA, which is an automated series of checks applied to the *training* data of the framework, and QC, which is automated checking of model *outputs*.

# Training

We trained a 2D nnU-Net on a subset of randomly selected cases from the NHS dataset (n=2793) using five-fold cross validation on the training set. Subsequently, nnU-Net was applied as an ensemble of the five models resulting from this cross validation.

# Validation and statistical analysis

We performed an internal validation of our tool using the remaining NHS cases that were not used for training (n=414). Subsequently, we performed an external validation in five additional datasets (n=6888) which included clinical CMR scans of patients with a range of diseases acquired at 10 international centres using 1.5 and 3T CMR scanners from all major vendors (Canon Medical Systems Corporation, Otawara, Tochigi, Japan; General Electric Healthcare, Chicago, Illinois, USA; Philips Healthcare, Best, the Netherlands; Siemens Healthineers, Erlangen, Germany). In this manuscript external database are considered to be those from different centres from the data used for training the model.

Dice scores were used to assess the agreement between the automated and manual segmentations. A Dice score of 0% indicates no agreement and a Dice score of 100% indicates perfect agreement. Distributions of Dice scores were tested for symmetry using D'Agostino's $K^2$ test and are reported using median (interquartile range) values. To provide more clinically meaningful validation, we used Bland-Altman plots to compare the LV and RV end-diastolic volume (EDV), end-systolic volume (ESV), ejection fraction (EF), and left ventricular mass (LVM) obtained using our method versus manual analysis, and calculated absolute errors (i.e., $|automated - ground\ truth|$) for each metric. See Supplemental Method H for more details of the computation of the cardiac biomarkers.

Mann-Whitney U tests were performed to compare both the Dice scores and the cardiac biomarkers between the NHS validation cases and the external validation datasets. Finally, we used box plots to compare performance over major groups of cardiac diseases and scanner vendors for CMR cases. Wilcoxon signed-ranked tests were applied to determine whether there were significant errors in biomarker estimation for the different disease and CMR vendor groups. *Post hoc* Bonferroni correction for multiple comparisons was performed for all statistical tests.

We additionally compared the segmentation performance of our tool against human inter-observer variability. To do so, a set of 50 subjects from the NHS validation dataset was randomly selected and each subject was analysed by three level 3 accredited CMR clinicians (O1, O2, O3) independently. The difference of clinical measurements was evaluated between each pair of observers (O1 vs O2, O2 vs O3, O3 vs O1) and between manual and automated analysis (see Supplemental Table 5).

## Comparison to other state-of-the-art methods

We compare the proposed method with the original nnU-Net AI trained model on the M&Ms dataset [10, 15] and tested on the internal NHS validation dataset. For validation, we computed the Dice scores between the automated and manual segmentations.

# Results

## Data identification

For the two clinical databases (internal NHS dataset and the Duke database), our image classification framework [12] correctly detected the cine SAX sequence in 100% of cases.

## Dice

For the internal NHS dataset, median Dice scores between the automated and manual segmentations were 94.3% for the LV blood pool (LVBP), 85.5% for the LV myocardium (MYO), and 90.8% for the RV blood pool (RVBP). For the external datasets, median Dice scores of the automated compared to manual segmentations were >91.3% for the LVBP, >83.0% for the MYO, and >87.4% for the RVBP. Dice scores for each dataset and cardiac label are shown in Table 2. Additionally, Dice scores for each scanner model and cardiac label are shown in Supplemental Table 6.

## Cardiac biomarkers

A Bland-Altman analysis of the differences between cardiac biomarkers derived from the manual and automated segmentations is shown in Figure 2. For the internal NHS dataset, median absolute errors in cardiac biomarkers were 6.7mL for LVEDV, 6.3mL for LVESV, 3.4% for LVEF, 8.9g for LVM, 8.5mL for RVEDV, 6.4mL for RVESV, and 4.2% for RVEF.

For the external datasets, median absolute errors in cardiac biomarkers were <7.3mL for LVEDV, <8.4mL for LVESV, <5.3% for LVEF, <13.3g for LVM, <9.2mL for RVEDV, <8.9mL for RVESV and <5.9% for RVEF. There was no significant bias for cardiac volumes or ejection fraction for the internal or external databases.

Table 3 shows, for each dataset, the ground truth values (first row) and the median (interquartile range) absolute errors (second row) of the automated analysis. There are statistically significant differences in performance between the internal NHS validation cases and the external validation datasets. The box plots in Figures 3 and 4 show the errors between ground truth and automated cardiac biomarkers grouped by cardiac disease and scanner type, with the largest errors found in healthy (NOR) subjects and Siemens scanners, respectively. Similar box plots grouped by magnetic field strength are shown in Supplemental Figure 3.

To validate the impact of our automated checking of ground truth segmentations ($QA^{gt}$), Supplemental Table 7 shows a comparison for the internal NHS validation dataset between the proposed method with and without $QA^{gt}$. The results demonstrate the importance of performing stringent automated data quality assessments on training databases. Supplemental Table 7 also shows results for a nnU-Net model trained only using the M&Ms database, demonstrating the necessity to train CMR segmentation models on large and heterogeneous databases.

## Quality control

From the test set segmentations of the CMR segmentation model, 607 cases out of 7302 were flagged, of which 18 were automatically adapted. The remaining cases that were flagged were not adapted and all were included in the segmentation and biomarker estimation results presented above. For the internal database, 6.59 % of cases were flagged, and on average, for the external datasets, 7-15% of cases were flagged (see Supplemental Table 4). To validate the impact of our automated QC of model outputs, Supplemental Table 8 and Supplemental Table 9 show the improvement in Dice scores and errors in ventricular volume quantification with and without post-analysis QC. For all labels (LV blood pool, LV myocardium and RV blood pool) Dice scores improved with post-analysis QC. The error in ventricular volumes also decreased. The largest decreases were seen for LVM and RVESV.

# Discussion

In this paper, we have proposed and validated the first start-to-end pipeline for fully automated analysis of large, unstructured clinical and research databases of CMR scans. This work addresses two majors but often-overlooked challenges in the exploitation of large CMR databases. First, data are often stored in an inconsistent and/or unstructured manner (e.g. different file format/structure, different segmentation protocols). Second, ground truth data often contain errors. We have proposed AI tools that address these challenges and hence enable for the first time fully automated processing of large databases of SAX CMR scans as well as flagging and potentially correcting error cases. Furthermore, our framework incorporates automated QC of its outputs, enabling the AI to “know when it has failed”, which is an important characteristic for full automation of routine but laborious clinical tasks.

We have validated our tool on a total of 7,293 CMR scans, including on two large clinical datasets containing routine CMR exams (n=1270 and n=414) and on four external research datasets (n=4787, n=345, n=345, and n=132). Through this extensive validation, we have shown that our tool achieves human-level accuracy for SAX cine CMR segmentation across a wide range of diseases, vendors, and clinical imaging protocols. Our method yielded median Dice scores of >91%, >83%, and >87% for the LVBP, LVM, and RVBP, respectively, translating into median *absolute errors* in cardiac biomarkers of <8.4mL for the LV, <9.2mL for the RV <13.3g for the LVM, and <5.9% for EF across all datasets. Our results (Table 3) did show a statistically significant difference in the accuracy of some clinical biomarkers between the internal and external validation databases. However, the limits of agreement between manual assessment and our AI tool for cardiac volumes was similar to the inter-observer variability observed in analysis of a sample of our own data by three independent clinical experts (see Supplemental Table 5) and inter-observer variability values reported in the literature [7, 16–18]. Note that it is to be expected that our method would not surpass these limits of agreement between observers, as our ground-truth data was segmented by a range of different experts from different institutions.

We examined the performance of our method over different phenotypes of cardiac disease (see Figure 2). Performance was good across different disease groups, including the spectrum of cardiomyopathic diseases, in addition to several congenital cases such as pulmonary arterial hypertension and tetralogy of Fallot (see Figure 5 for examples). Although the errors were significantly different from zero in some groups, they were within the ranges of inter-observer variability in manual CMR analysis [7, 16–18], and are therefore similar to those experienced in clinical practice; the worst-performing phenotype (healthy subjects – NOR category in Figure 2) had median absolute errors of <10mL for volumes and <5% for EF. Most healthy subjects came from the UK Biobank dataset, whose segmentations were obtained using a stringent standardised operating procedure whereby each analyst was specifically trained to limit inter-observer variability, thus leading to a systematic segmentation style. This could explain why this group had the highest bias in our analysis. The training dataset (NHS) is likely to reflect the more typical level of inter-observer variability among CMR cardiologists, which also exists in all validation datasets except for the UK Biobank. As an adult congenital heart disease (CHD) list is part of the routine CMR service of the NHS hospitals, some (but a limited number of) CHD cases were included in training and testing. This included patients with (repaired) tetralogy of Fallot, transposition of the great arteries, bicuspid aortic valves and pulmonary valve diseases. Patients with single-ventricle circulation were excluded. An analysis of congenital disease phenotypes is beyond the scope of the current work, but will be subject of a future study. Overall, these results show that our tool performs well for a typical (non-congenital) CMR list.

Our method also performed well across all scanner types (see Figure 4), including those not seen during training (Canon and GE). The largest errors were found in Siemens data, but this error was again within inter-observer variability in CMR analysis [17, 18]. Again, this is likely explained by the systematic difference in segmentation style between the UK Biobank data – which was acquired using a single Siemens scanner model – and the NHS data.

We deliberately trained the segmentation model on one clinical dataset (the NHS dataset) alone, without using the full variety of scanner vendors and protocols in the external datasets. This reflects the real-world challenge of application of AI tools in previously unseen data, due to protocol and scanner updates. The good domain generalisability of our method to unseen data in the validation experiments demonstrates the strength of our framework in generalising to data unseen during training.

A number of previous studies have evaluated AI segmentation models on CMR data, although typically on smaller/non-existent and/or less variable external validation sets. Nevertheless, the performance of our method is comparable to previously published algorithms, e.g. see [2, 4–6, 9, 19–21] and the references therein. Furthermore, we have compared the proposed method with the original nnU-Net AI trained model on the M&Ms dataset (Supplemental Table 7) and show an improved performance.

## Novelty of our method

Over the last years many research and commercial solutions have been proposed for automated segmentation of LV and RV volumes from SAX cine CMR sequences. However, these methods are only suitable for analysis of data at the point of acquisition (where manual QC can be routinely performed) or of highly structured databases such as the UK Biobank. Such databases are relatively uncommon, but our AI-based pipeline enables fully automated analysis of a wider range of clinical databases, which are typically unstructured and/or contain errors. Furthermore, most existing automated CMR analysis methods have been developed and validated on a single highly controlled dataset, with limited external validation [2, 6]. One recent study used multi-centre, multi-vendor data for training [4]; however, it did not include Canon scanners and only validated externally using a single-centre, single-scanner dataset. AI methods tend to not be domain agnostic and can generalise poorly to other domains [22]. This is an issue not only in research, but also affects algorithms already being deployed in commercial CMR analysis software packages [16]. Therefore, extensive external validation of new AI algorithms in large, heterogeneous datasets that include a range of diseases is an important step to perform prior to clinical translation. In this study we have aggregated CMR data from two UK hospitals, one US hospital, and several external research datasets to perform such a validation.

We also stratified our models' validation experiments over groups of patients with different disease phenotypes. This is another important aspect of validation of clinical AI tools, as errors in segmentation algorithms might impact some disease phenotypes more than others.

Altogether the new contributions in this paper, combined with our previous developments for automated detection of cine CMR views from full CMR scans [12], and post-analysis QC of the output parameters for physiological feasibility [2], show that our framework is a robust tool for automated analysis of large, unstructured databases of CMR scans. It can automatically identify CMR scans from a file/folder structure (even if this structure includes other data), detect target cine CMR sequences for analysis, analyse these images and provide parameters of biventricular function accurately and robustly for all major disease groups, scanner vendors and imaging protocols, while flagging cases with potential errors in data.
Finally, we are in the process of making this tool available as an easy-to-use web application which will be accessible for external researchers.

## Quality-controlled AI

Most commercial cardiac analysis software has already implemented AI algorithms. For example, a recent study [7] compared the performances of three different commercial solutions (CardioAI (Arterys), CVI42 (Circle Cardiovascular Imaging) and SuiteHeart (Neosoft) in a cohort of 200 ischemic heart disease patients. Although they show excellent agreements with manual annotation, most of these algorithms are still not optimised to generalise to out-of-distribution data, and suffer from reduced performance when applied to other datasets [7, 22]. Therefore, these tools require continuous clinician oversight, which is sufficient and desirable for prospective clinical reporting. However, in the current era of big data, AI tools that can analyse large (retrospective) datasets or registries robustly are essential for developing clinical research. In the current work, we detect and automatically correct some common errors in segmentations (e.g. additional segmented regions outside of the heart area). More complex distortions were not addressed, but together with our previously developed post-analysis QC steps of the obtained ventricular volumetrics [2], the overall QC process can be used to flag potential errors during automated analysis to clinicians for review when the tool is implemented for use. This aids a trustworthy and transparent system.

The beneficial effect of post-analysis QC and automated adaption is demonstrated by the improvements in Dice scores and volume errors when comparing the AI algorithms' output with and without QC shown in Supplemental Tables 8 and 9. Dice scores and the absolute error in volume quantification between AI and manual assessment improved for all cardiac metrics. The largest improvements were found for LVM and RV parameters. LVM and RV anatomy are known to be the most challenging tasks for AI algorithms, and these improvements are therefore clinically relevant.

The post-analysis QC flagged 6-15% of cases from the output of the segmentation model in the different test datasets (see Supplemental Table 4). The highest rates were in the ACDC and M&Ms databases, and nearly all flags related to an SV difference >25%. This was expected, as many of the cardiomyopathy patients in these datasets suffer from valvar regurgitation. Flagged cases therefore do not always reflect errors in the results, but aid to highlight challenging / unusual cases during large database processing for clinicians for review.

Other techniques have been proposed in the literature for QC of segmentation model outputs, and these could also be incorporated into our framework if they brought added value. For example, Robinson et al [28] proposed to use an atlas registration-based approach to estimate segmentation quality. Uncertainty-based approaches to segmentation QC have been proposed by Puyol Anton et al [3] and Arega et al [23], both for segmentation of T1 mapping CMR images.

## The importance of ground truth data quality

The quality of the ground truth segmentations is important when developing and evaluating AI methods, but most existing works assume that ground truth quality is high without checking. Therefore, we performed automated data quality assessments ($QA^{gt}$) on all datasets used for training and validation. The flagging of potential erroneous segmentations was fully automated and followed by a manual inspection by expert cardiologists if the error could not be corrected automatically. Through this $QA^{gt}$ we excluded a number of erroneous ground truth segmentations (e.g., a partial segmentation that was stored halfway through the manual analysis and was never completed by the clinician) from the clinical datasets. However, perhaps surprisingly a number of cases from the external validation datasets (i.e., ACDC, M&Ms and M&Ms-2) were also excluded, as segmentation quality of these cases did not adhere to the clinical standards for ventricular segmentation for volume quantification published by the European and American CMR societies [24, 25]. The majority of these errors consisted of the absence of the basal slices in the short-axis cine acquisition (see examples of erroneous ground truth segmentations in Figure 6). Since the basal slices are challenging for segmentation algorithms, missing basal slices have direct implications on validation results. Another common challenge was the segmentation of the RV base. Segmentations did not always continue until the pulmonary valve as is recommended in the standards of segmentation by the European and American CMR societies [24, 25] (see Figure 6.b which shows a ground truth segmentation missing the top basal slice).

Our $QA^{gt}$ of the ground truth segmentations from the external validation datasets led to the exclusion of 1.4% of cases (see Supplemental Table 2). While this number is relatively small, we argue that assuring high standards for quality of ground truth data is important when training AI algorithms. Supplemental Figure 4 shows examples of cases flagged by $QA^{gt}$ for the external validation datasets and Supplemental Method C includes the list of all cases excluded from the online available databases (ACDC, M&Ms and M&Ms2). Supplemental Table 7 shows that $QA^{gt}$ improves the robustness of the resulting trained segmentation model. Note that, for completeness and to allow evaluation against previously published results, we have included the performance of our method on the full original external validation datasets (i.e. without QA) in Supplemental Tables 8 and 9.

In our work we have proposed to use a set of heuristic rules for QA of segmentations. However, in principle other QA approaches could be substituted within our framework. For example, the recent literature has proposed a range of approaches for image QA, including methods based on hand-crafted quality metrics [26] and deep learning-based methods [27] for detecting predefined types of image artifacts. Methods proposed for QA of segmentations include the supervised approach of Fournel *et al.* [28], who proposed to directly predict 2D and 3D Dice scores from segmentation/image pairs and the unsupervised approach of Galati and Zuluaga [29], who used the reconstruction error of a convolutional autoencoder trained on ground truth segmentations as a surrogate measure of segmentation quality. All of these approaches could in principle be added to our framework and in the future, we will investigate if they bring added benefit over our current approach.

## Dealing with papillary muscles

Different protocols are used when segmenting short-axis cine CMR images. Some CMR departments include papillary muscles in the LV and RV blood pools, while others exclude them. This depends both on clinician preference and on the segmentation technique used (e.g., drawing myocardial borders manually or semi-automatically using region growing) [25]. Our tool was originally trained to include papillary muscles in the ventricular blood pools. However, papillary muscles were excluded from the segmentations in the Duke external validation dataset. Therefore, we developed and applied an automatic image thresholding technique – based on Otsu's method – to apply to the segmented blood pools to obtain segmentations that exclude the papillary muscles. This not only allowed external validation on the Duke dataset but also allows users of our tool to choose their preferred segmentation strategy.

## Limitations

Although we trained our segmentation algorithm on a large clinical dataset (n=2793) and showed good generalisation to out-of-distribution data, the validation data did not include all variations of data in clinical practice. In particular complex congenital heart diseases were absent from our training data. Similarly, our five external validation datasets are not all-encompassing. In the future, we aim to deploy our method in other centres to perform additional validation steps throughout deployment to ensure the tool's robustness and to improve its performance. Furthermore, the current version of the tool can only be used to analyse SAX CMR scans. Future work will extend the tool to be used with long-axis cine CMR.

# Conclusions

We have developed a robust start-to-end AI-based tool for quality-controlled, automated analysis of short-axis CMR scans. We implemented a state-of-the-art AI method that significantly reduces the performance drop for CMR images not seen during training. We validated our tool using over 7,000 CMR cases from multiple centres and countries and showed that our method yields human-level accuracy for LV and RV segmentations for all major CMR scanner vendors, and for a wide range of cardiac disease phenotypes and acquisition protocols.

**ABBREVIATIONS AND ACRONYMS:**

AI = artificial intelligence

CMR = cardiac magnetic resonance

ED = end-diastole/diastolic

EDV = end-diastolic volume

EF = ejection fraction

ES = end-systole/systolic

ESV = end-systolic volume

LV = left ventricle/ventricular

LVBP = left ventricular blood pool

LVM = left ventricular mass

MYO = left ventricular myocardium

QA = quality assessment

QC = quality control

RV = right ventricle/ventricular

RVBP = right ventricular blood pool

**DECLARATIONS:**

**ETHIC APPROVALS**

Our study complies with the Declaration of Helsinki. Ethical approval for use of the NHS dataset was obtained from the London Dulwich Research ethics committee (ID: 19/LO/1957). The Duke data was collected as part of an ethically approved study with title 'Deidentified Cardiovascular Magnetic Resonance Images' with Duke University Research Ethics Committee reference Pro00056051 and was fully de-identified and transferred to King's College London via a data use agreement approved by both institutions. Use of the UK Biobank dataset was covered by the ethical approval from the NHS National Research Ethics Service on 17th June 2011 (Ref 11/NW/0382) and extended on 18 June 2021 (Ref 21/NW/0157) with written informed consent obtained from all participants. The other datasets are available online, under previous approval from the various host institute ethical committees.

**CONSENT FOR PUBLICATION**

Not applicable.

## AVAILABILITY OF DATA AND MATERIALS

Some of the datasets presented in this study can be found in online repositories. The names of the repository/repositories and accession number(s) can be found below: The UK Biobank data set is publicly available for approved research projects from https://www.ukbiobank.ac.uk/. The ACDC, M&Ms and M&Ms2 data sets are publicly available to approved participants from https://acdc.creatis.insa-lyon.fr/description/databases.html, https://www.ub.edu/mnms/ and https://www.ub.edu/mnms-2/.

The NHS and Duke data sets cannot be made publicly available due to restricted access under hospital ethics and because informed consent from participants did not cover public deposition of data.

## COMPETING INTERESTS

SEP provided consultancy to Circle Cardiovascular Imaging, Inc., Calgary, Alberta, Canada. RMJ is an employee of Intelerad Medical Systems Inc., Montreal, Canada. The remaining authors declare that the research was conducted without any commercial or financial relationships that could be construed as a potential conflict of interest.

## FUNDING:

This work was supported by the EPSRC (EP/R005516/1) and the Advancing Impact Award scheme of the Impact Acceleration Account at King's College London. SEP, RR, and AK acknowledge funding from the EPSRC through the SmartHeart Programme grant (EP/P001009/1). JM-H, RR, AK, BR, and EP-A acknowledge support from the Wellcome/EPSRC Centre for Medical Engineering at King's College London (WT 203148/Z/16/Z), the National Institute for Health Research (NIHR) Cardiovascular MedTech Co-operative award to the Guy's and St Thomas' NHS Foundation Trust, and the NIHR comprehensive Biomedical Research Centre award to Guy's & St Thomas' NHS Foundation Trust in partnership with King's College London. SEP acknowledges the British Heart Foundation for funding the manual analysis to create a cardiovascular magnetic resonance imaging reference standard for the UK Biobank imaging resource in 5,000 CMR scans (www.bhf.org.uk; PG/14/89/31194). SEP acknowledges support from the NIHR Biomedical Research Centre at Barts. SEP has received funding from the European Union's Horizon 2020 Research and Innovation Programme under grant agreement No 825903 (euCanSHare project). SEP also acknowledges support from the CAP-AI Programme, London's First AI Enabling Programme focused on stimulating growth in the capital's AI Sector. CAP-AI was led by Capital Enterprise in partnership with Barts Health NHS Trust and Digital Catapult and was funded by the European Regional Development Fund and Barts Charity. SEP acknowledges support from the Health Data Research UK, an initiative funded by UK Research and Innovation, the Department of Health and Social Care (England) and the devolved administrations, and leading medical research charities. The views expressed are those of the author(s) and not necessarily those of the EPSRC, the NHS, the NIHR, the Department of Health, or the British Heart Foundation. For the purpose of open access, the author has applied for a CC BY public copyright licence for any Author Accepted Manuscript version arising from this submission. This research has been conducted using the UK Biobank Resource (application 17806). The UK Biobank data are available for approved projects from https://www.ukbiobank.ac.uk/.

**AUTHORS' CONTRIBUTIONS**

JM-H, BR, and EP-A designed, developed, and tested the method and analysed the data. JM-H, RR, APK, BR, and EP-A conceived the study. CA, VV, MR, LK, RKJ, RMJ, SEP, and BR provided the manual segmentations used for the development and testing of the method. JM-H, BR, and EP-A drafted the work and CA, VV, MR, LK, RJK, RMJ, SEP, RR, and APK revised it critically for important intellectual content. RR and APK were part of the supervision of JM-H during method development.

**ACKNOWLEDGEMENTS**

This research has been conducted using the UK Biobank Resource under Application Number 17806. The authors wish to thank all participants and staff from UK Biobank, NHS, ACDC, M&Ms and M&Ms2.

# References

1. Von Knobelsdorff-Brenkenhoff F, Pilz G, Schulz-Menger J. Representation of cardiovascular magnetic resonance in the AHA / ACC guidelines. J Cardiovasc Magn Reson. 2017;19:1–21.

2. Ruijsink B, Puyol-Antón E, Oksuz I, Sinclair M, Bai W, Schnabel JA, et al. Fully Automated, Quality-Controlled Cardiac Analysis From CMR. JACC Cardiovasc Imaging. 2020;13:684–95.

3. Puyol-Antón E, Ruijsink B, Baumgartner CF, Masci P-G, Sinclair M, Konukoglu E, et al. Automated quantification of myocardial tissue characteristics from native T1 mapping using neural networks with uncertainty-based quality-control. J Cardiovasc Magn Reson. 2020;22:60.

4. Davies RH, Augusto JB, Bhuva A, Xue H, Treibel TA, Ye Y, et al. Precision measurement of cardiac structure and function in cardiovascular magnetic resonance using machine learning. J Cardiovasc Magn Reson. 2022;24:1–11.
5. Fadil H, Totman JJ, Hausenloy DJ, Ho HH, Joseph P, Low AFH, et al. A deep learning pipeline for automatic analysis of multi-scan cardiovascular magnetic resonance. J Cardiovasc Magn Reson. 2021;23:1–13.
6. Bai W, Sinclair M, Tarroni G, Oktay O, Rajchl M, Vaillant G, et al. Automated cardiovascular magnetic resonance image analysis with fully convolutional networks. J Cardiovasc Magn Reson. 2018;20:65.
7. Wang S, Patel H, Miller T, Ameyaw K, Narang A, Chauhan D, et al. AI Based CMR Assessment of Biventricular Function: Clinical Significance of Intervendor Variability and Measurement Errors. JACC Cardiovasc Imaging. 2022;15:413–27.
8. Petersen SE, Matthews PM, Francis JM, Robson MD, Zemrak F, Boubertakh R, et al. UK Biobank's cardiovascular magnetic resonance protocol. J Cardiovasc Magn Reson. 2015;18:8.
9. Bernard O, Lalande A, Zotti C, Cervenansky F, Yang X, Heng P-A, et al. Deep Learning Techniques for Automatic MRI Cardiac Multi-Structures Segmentation and Diagnosis: Is the Problem Solved? IEEE Trans Med Imaging. 2018;37:2514–25.
10. Campello VM, Gkontra P, Izquierdo C, Martin-Isla C, Sojoudi A, Full PM, et al. Multi-Centre, Multi-Vendor and Multi-Disease Cardiac Segmentation: The MMs Challenge. IEEE Trans Med Imaging. 2021;40:3543–54.
11. Martín-Isla C, K. L. Multi-Disease, Multi-View & Multi-Center Right Ventricular Segmentation in Cardiac MRI (M&Ms-2).
12. Vergani V, Razavi R, Puyol-Antón E, Ruijsink B. Deep Learning for Classification and Selection of Cine CMR Images to Achieve Fully Automated Quality-Controlled CMR Analysis From Scanner to Report. Front Cardiovasc Med. 2021;8 October.
13. Isensee F, Jaeger PF, Kohl SAA, Petersen J, Maier-Hein KH. nnU-Net: a self-configuring method for deep learning-based biomedical image segmentation. Nat Methods. 2021;18:203–11.
14. Otsu N. A Threshold Selection Method from Gray-Level Histograms. IEEE Trans Syst Man Cybern. 1979;C:62–6.
15. Full PM, Isensee F, Jäger PF, Maier-Hein K. Studying robustness of semantic segmentation under domain shift in cardiac MRI. Stat Atlases Comput Model Hear M&Ms EMIDEC Challenges 11th Int Work STACOM 2020, Held Conjunction with MICCAI 2020, Lima, Peru. 2020;:238–49.
16. Petersen SE, Aung N, Sanghvi MM, Zemrak F, Fung K, Paiva JM, et al. Reference

ranges for cardiac structure and function using cardiovascular magnetic resonance ( CMR ) in Caucasians from the UK Biobank population cohort. J Cardiovasc Magn Reson. 2017;19:1–19.
17. Childs H, Ma L, Ma M, Clarke J, Cocker M, Green J, et al. Comparison of long and short axis quantification of left ventricular volume parameters by cardiovascular magnetic resonance, with ex-vivo validation. J Cardiovasc Magn Reson. 2011;13:1–9.
18. Luijnenburg SE, Robbers-Visser D, Moelker A, Vliegen HW, Mulder BJM, Helbing WA. Intra-observer and interobserver variability of biventricular function, volumes and mass in patients with congenital heart disease measured by CMR imaging. Int J Cardiovasc Imaging. 2010;26:57–64.
19. Penso M, Moccia S, Scafuri S, Muscogiuri G, Pontone G, Pepi M, et al. Automated left and right ventricular chamber segmentation in cardiac magnetic resonance images using dense fully convolutional neural network. Comput Methods Programs Biomed. 2021;204:106059.
20. Budai A, Suhai FI, Csorba K, Toth A, Szabo L, Vago H, et al. Fully automatic segmentation of right and left ventricle on short-axis cardiac MRI images. Comput Med Imaging Graph. 2020;85:101786.
21. Suinesiaputra A, Mauger CA, Ambale-Venkatesh B, Bluemke DA, Dam Gade J, Gilbert K, et al. Deep Learning Analysis of Cardiac MRI in Legacy Datasets: Multi-Ethnic Study of Atherosclerosis. Front Cardiovasc Med. 2022;8.
22. Leiner T, Rueckert D, Suinesiaputra A, Baeßler B, Nezafat R, Išgum I, et al. Machine learning in cardiovascular magnetic resonance: Basic concepts and applications. J Cardiovasc Magn Reson. 2019;21:1–14.
23. Arega TW, Bricq S, Legrand F, Jacquier A, Lalande A, Meriaudeau F. Automatic uncertainty-based quality controlled T1 mapping and ECV analysis from native and post-contrast cardiac T1 mapping images using Bayesian vision transformer. Med Image Anal. 2023;86:102773.
24. Petersen SE, Khanji MY, Plein S, Lancellotti P, Bucciarelli-Ducci C. European Association of Cardiovascular Imaging expert consensus paper: A comprehensive review of cardiovascular magnetic resonance normal values of cardiac chamber size and aortic root in adults and recommendations for grading severity. Eur Heart J Cardiovasc Imaging. 2019;20:1321–31.
25. Schulz-Menger J, Bluemke DA, Bremerich J, Flamm SD, Fogel MA, Friedrich MG, et al. Standardized image interpretation and post-processing in cardiovascular magnetic resonance - 2020 update: Society for Cardiovascular Magnetic Resonance (SCMR): Board of Trustees Task Force on Standardized Post-Processing. J Cardiovasc Magn Reson.

2020;22:1–22.
26. Tarroni G, Bai W, Oktay O, Schuh A, Suzuki H, Glocker B, et al. Large-scale Quality Control of Cardiac Imaging in Population Studies: Application to UK Biobank. Sci Rep. 2020;10:2408.
27. Oksuz, Ilkay and Ruijsink, Bram and Puyol-Antón, Esther and Sinclair, Matthew and Rueckert, Daniel and Schnabel, Julia A and King AP, Oksuz I, Ruijsink B, Puyol-Anton E, Sinclair M, Rueckert D, et al. Automatic left ventricular outflow tract classification for accurate cardiac MR planning. In: 2018 IEEE 15th International Symposium on Biomedical Imaging (ISBI 2018). IEEE; 2018. p. 462–5.
28. Fournel J, Bartoli A, Bendahan D, Guye M, Bernard M, Rauseo E, et al. Medical image segmentation automatic quality control: A multi-dimensional approach. Med Image Anal. 2021;74:102213.
29. Galati F, Zuluaga MA. Efficient Model Monitoring for Quality Control in Cardiac Image Segmentation. Functional Imaging and Modeling of the Heart: 11th International Conference, FIMH 2021, Stanford, CA, USA. Springer International Publishing; 2021. p. 101–11.

**FIGURE TITLES AND LEGENDS:**

**Figure 1 – AI-based, quality-controlled CMR analysis tool:** training, validation, and CMR analysis tool outline. $QA^{gt}$: ground truth segmentation data quality assessment, QC: quality control.

**Figure 2 – Bland-Altman analysis of cardiac volume, ejection fraction, and mass:** Cardiac biomarkers derived from manual and automated segmentations were compared for all validation cases. The thick line depicts the mean bias between the automated and manual analysis. The top and bottom dotted lines correspond to +1.96 and -1.96 standard deviations from the mean bias, respectively. The Pearson's Correlation Coefficients (R) between our method and the manual analysis (and the corresponding p-values) are indicated for each cardiac biomarker. LVEDV: left ventricular end-diastolic volume; LVESV: left ventricular end-systolic volume; LVEF: left ventricular ejection fraction; LVM: left ventricular mass; RVEDV: right ventricular end-diastolic volume; RVEF: right ventricular ejection fraction; and RVESV: right ventricular end-systolic volume.

**Figure 3 – Box plots of manually and automatically derived cardiac biomarkers for each disease group:** Statistical differences from zero were assessed using Wilcoxon signed rank tests. Pairwise post hoc testing was performed using Bonferroni correction for multiple comparisons. Asterisks indicate statistically significant differences from zero for each group after correction (five tests), where * = $p < 0.01/5$, ** = $p < 0.001/5$, *** = $p < 0.0001/5$. CHD: Congenital Heart Disease; DCM: Dilated Cardiomyopathy; IHD: Ischemic Heart Disease; LVEDV: left ventricular end-diastolic volume; LVESV: left ventricular end-systolic volume; LVEF: left ventricular ejection fraction; LVM: left ventricular mass; NOR: Normal Cardiac Anatomy and Function; Other: other cardiac diseases; RVEDV: right ventricular end-diastolic volume; RVEF: right ventricular ejection fraction; and RVESV: right ventricular end-systolic volume.

**Figure 4 – Box plots of manually and automatically derived cardiac biomarkers for each scanner group:** Statistical differences from zero were assessed using Wilcoxon signed rank tests. Pairwise post hoc testing was performed using Bonferroni correction for multiple comparisons. Asterisks indicate statistically significant differences from zero for each group after correction (four tests), where * = $p < 0.01/4$, ** = $p < 0.001/4$, *** = $p < 0.0001/4$. Abbreviations as in Figure 2.

**Figure 5 – Examples of automated segmentations from different disease groups:** The middle slice of each automated segmentation is shown in ED and ES. Cardiac labels: LV blood pool (red), LV myocardium (yellow) and RV blood pool (blue).

**Figure 6 – Examples of erroneous ground truth segmentations identified during manual $QA^{gt}$:** Cardiac labels: LV blood pool (red), LV myocardium (yellow) and RV blood pool (blue). a) Image in ED: note that the LV myocardium is segmented, but the LV blood pool segmentation is absent and that the RV segmentation is labelled as myocardium (yellow); b) top slice of the cine stack in ED: the basal part of the heart is not included in the cine SAX stack; c) image in ED: note the unusual LV structure that was segmented and the absence of an RV segmentation; d) image in ES; note the absence of LV and RV segmentations, while myocardium is present for both.

**TABLES:**

***Table 1 – Internal and external validation dataset characteristics***

| Dataset | Country | Centre | Scanner vendor | Scanner model | Disease |
|---|---|---|---|---|---|
| NHS (n=414) | UK | Guy's and St Thomas' NHS Foundation Trust | Philips | Achieva 1.5T/3.0T | CHD (n=13) DCM (n=29) IHD (n=15) NOR (n=39) Other (n=50) N/A (n=268) |
| | | | | Ingenia 1.5T | |
| | | | Siemens | Aera 1.5T Biograph mMR | |
| Duke (n=1319) | US | Duke University Hospital | Siemens | Avanto 1.5T | N/A (n=1319) |
| | | | | Sola 1.5T | |
| | | | | Verio 3.0T | |
| | | | | Vida 3.0T | |
| UKBB (n=4872) | UK | 4 centres | Siemens | Aera 1.5T | NOR (n=4872) |
| ACDC (n=150) | France | Centre Hospitalier Universitaire Dijon Bourgogne | Siemens | Aera 1.5T | ARV (n=30) DCM (n=30) HCM (n=30) IHD (n=30) NOR (n=30) |
| | | | | Trio 3.0T | |
| M&Ms (n=375) | Spain | Clínica Creu Blanca | Canon | Orian 1.5T | AHS (n=3) ARV (n=16) DCM (n=51) HCM (n=103) HHD (n=25) IHD (n=8) LVNC (n=4) NOR (n=125) Other (n=40) |
| | | Hospital Universitari Dexeus | GE | Excite 1.5T | |
| | | Clínica Sagrada Familia | Philips | Achieva 1.5T | |
| | | Hospital Vall d'Hebron | Siemens | Avanto 1.5T | |
| | Germany | Universitätsklinikum Hamburg-Eppendorf | Philips | Achieva 1.5T | |
| | Canada | McGill University Health Centre | Siemens | Skyra 3.0T | |
| M&Ms-2 (n=360) | Spain | Hospital Universitari Dexeus | GE | Excite 1.5T | ARR (n=35) DLV (n=60)* |
| | | | | Explorer 1.5T | |
| | | | | HDxt 1.5T/3.0T | DRV (n=30)<br>HCM (n=60)*<br>CIA (n=35)<br>NOR (n=75)*<br>FALL (n=35)<br>TRI (n=30) |
| | | Clínica Sagrada Familia | Philips | Achieva 1.5T | |
| | | Hospital Vall d'Hebron | Siemens | Avanto 1.5T | |
| | | | | Symphony 1.5T | |
| | | | | Trio 3.0T | |

Dataset names, countries, centres, scanner vendors and models, and cardiac diseases. For the NHS database, the CHD cases consisted of patients with bicuspid aortic valve (N=6), (repaired) tetralogy of Fallot and equivalent (N=2), atrial septal defects (N=2), aortic and pulmonary regurgitation (N=2), and repaired transposition of great arteries (N=1). *These subjects were also included in M&Ms. AHS: athletic heart syndrome; ARR: congenital arrhythmogenesis; ARV: abnormal right ventricle; CIA: interatrial communication; CHD: congenital heart disease; DCM: dilated cardiomyopathy; DLV: dilated left ventricle; DRV: dilated right ventricle; FALL: tetralogy of fallot; HCM: hypertrophic cardiomyopathy; HHD: hypertensive heart disease; IHD: ischemic heart disease; LVNC: left ventricular non-compaction; NOR: normal cardiac anatomy and function; TRI: tricuspid regurgitation; N/A: Disease information not available for these patients; UKBB: UK Biobank.

***Table 2 – Dice scores of automated vs manual segmentations***

| Dataset | LVBP [%] | MYO [%] | RVBP [%] | Average [%] |
|---|---|---|---|---|
| NHS | 94.3 (4.0) | 85.5 (4.4) | 90.8 (5.4) | 91.3 (7.5) |
| Duke | 91.3 (6.5)*** | 83.0 (4.8)*** | 89.3 (6.8)*** | 89.2 (8.3)*** |
| UKBB | 91.8 (5.9)*** | 83.0 (6.5)*** | 87.8 (7.8)*** | 87.4 (8.9)*** |
| ACDC | 95.5 (3.9) | 87.4 (3.5)*** | 91.8 (7.0)* | 90.6 (7.9) |
| M&Ms | 93.4 (5.4)*** | 85.4 (5.7) | 90.4 (6.8) | 89.4 (8.3)*** |
| M&Ms-2 | 94.6 (4.2) | 86.0 (5.6)* | 90.9 (6.7) | 90.3 (8.3)** |

Dice scores are shown as median (interquartile range) percentages for each validation dataset, including values per label and their average. Comparisons between the Dice scores of the NHS validation cases and the external validation datasets were performed using Mann-Whitney U tests. Pairwise *post hoc* testing was performed using Bonferroni correction for multiple comparisons. Asterisks indicate statistically significant differences for each label after correction (20 tests), where * = $p < 0.01/20$, ** = $p < 0.001/20$, *** = $p < 0.0001/20$. LVBP: LV blood pool, MYO: LV myocardium, RVBP: RV blood pool.

***Table 3 – Errors between cardiac biomarkers derived from automated and manual segmentations***

| | LV | | | | RV | | |
|---|---|---|---|---|---|---|---|
| **Dataset** | **EDV [mL]** | **ESV [mL]** | **EF [%]** | **LVM [g]** | **EDV [mL]** | **ESV [mL]** | **EF [%]** |
| NHS | 198.8 (75.3) | 108.9 (70.3) | 48.7 (14.0) | 116.5 (42.6) | 172.4 (52.4) | 84.6 (40.3) | 52.3 (9.8) |
| | 6.7 (13.5) | 6.3 (12.9) | 3.4 (4.3) | 8.9 (17.9) | 8.5 (15.8) | 6.4 (12.5) | 4.2 (5.4) |
| Duke† | 107.7 (46.6) | 50.7 (38.2) | 55.9 (14.8) | 107.4 (46.1) | 104.6 (41.4) | 52.6 (28.3) | 51.0 (11.2) |
| | 5.7 (11.0)*** | 5.0 (9.7) | 5.3 (7.1)*** | 13.3 (17.2)*** | 6.0 (11.8)*** | 4.5 (9.1)*** | 5.2 (6.8)** |
| UKBB | 149.0 (34.9) | 62.0 (20.7) | 58.8 (6.3) | 91.9 (25.4) | 155.6 (37.5) | 69.1 (22.7) | 56.1 (6.4) |
| | 7.3 (12.5)*** | 8.4 (9.6)*** | 4.7 (4.2)*** | 6.6 (12.8)** | 9.2 (14.7)*** | 8.9 (12.5)*** | 5.3 (5.9)*** |
| ACDC | 161.4 (67.6) | 94.0 (71.0) | 47.6 (18.8) | 126.7 (50.1) | 151.7 (51.2) | 86.1 (51.1) | 46.1 (18.1) |
| | 3.6 (5.8) | 5.3 (8.0)*** | 4.6 (5.1)*** | 6.8 (12.7)*** | 6.1 (11.2)*** | 5.7 (11.1)** | 5.6 (7.6) |
| M&Ms | 159.1 (61.1) | 73.0 (54.2) | 56.9 (13.7) | 115.9 (49.1) | 147.4 (49.9) | 69.5 (37.4) | 53.9 (12.2) |
| | 6.2 (13.4) | 5.1 (9.8)*** | 4.2 (5.4)*** | 8.2 (16.6)*** | 8.4 (16.9) | 5.9 (13.7)*** | 5.6 (7.3)*** |
| M&Ms-2 | 174.9 (61.6) | 89.6 (58.9) | 52.0 (13.7) | 112.0 (36.7) | 169.1 (56.5) | 89.5 (42.1) | 48.3 (12.8) |
| | 6.3 (9.6)*** | 5.5 (8.4)*** | 3.4 (4.2) | 7.3 (14.2)*** | 8.7 (16.5)*** | 8.3 (13.9)*** | 5.9 (7.0)*** |

For each validation dataset, the first row (highlighted in grey) reports the ground truth clinical measurements for each cardiac biomarker as median (interquartile range). The second row reports the median absolute errors (interquartile range) between cardiac biomarkers derived from the automated and ground truth segmentations. Comparisons between the errors of the NHS validation cases and the external validation datasets were performed using Mann-Whitney U tests. Pairwise *post hoc* testing was performed using Bonferroni correction for multiple comparisons. Asterisks indicate statistically significant differences for each biomarker after correction (35 tests), where * = $p < 0.01/35$, ** = $p < 0.001/35$, *** = $p < 0.0001/35$. † Lower ventricular volumes in the Duke dataset are due to the exclusion of papillary muscle from the ground truth segmentations. LV: left ventricle, RV: right ventricle, EDV: end-diastolic volume, ESV: end-systolic volume, EF: ejection fraction, LVM: left ventricular mass.

**TITLE**:
Large-scale, multi-centre, multi-disease validation of an AI clinical tool for cine CMR analysis

# Supplemental Appendix

## Supplemental Methods

## Supplemental tables

# Supplemental figures

# Supplemental Methods

## A. NHS dataset acquisition, pre-processing, and quality assessment

For the NHS dataset, clinical CMR scans and corresponding segmentations were automatically downloaded from the Guy's and St Thomas' NHS Foundation Trust PACS, sorted according to their acquisition sequence, and anonymised. They were then converted from a DICOM format to a NIFTI format as required by the tool, duplicate or empty ground truth segmentations were removed, end-diastolic and end-systolic frames were identified, and a thorough data quality assessment was applied to the ground truth segmentations from both frames. A summary of this process, including the number of cases discarded in each step, is provided in Supplemental Figure 1. The manual segmentations (i.e., left ventricular endocardium and myocardium and right ventricular endocardium) were acquired using the commercially available cvi42 CMR analysis software (Circle Cardiovascular Imaging Inc., Calgary, Alberta, Canada, version 5.10.1). For the NHS dataset the data quality assessment was independent to the automated quality assessment of ground truth segmentation performed on the external databases (sections B and C).

## B. Automated quality assessment of ground truth segmentations

We used statistical criteria together with clinical knowledge to establish the following criteria to identify potentially erroneous ground truth segmentations:

- containing unexpected labels - other than LVBP, MYO, or RVBP
- one of its ventricular volumes being equal to zero
- disagreement between image and segmentation metadata (e.g., different voxel size or orientation)
- containing outliers (i.e., voxels disconnected from the main segmentation) representing more than 10% of the total number of segmented voxels - this step was first applied to a combination of all segmented labels and then to each individual label
- any label appearing in less than 30% of segmented slices
- presence of non-segmented slices between segmented slices for any label
- negative or zero LV or RV stroke volume
- LV and RV stroke volume differences larger than 25%
- Missing basal slices on the acquisition or for segmenting

Finally, outliers representing less than 10% of the segmented voxels were automatically removed, keeping only the largest connected component for each label.

## C. Manual quality assessment of ground truth segmentations

The potentially erroneous ground truth segmentations flagged from section B from the Duke, ACDC, M&Ms, and M&Ms-2 datasets underwent a manual $QA^{gt}$ that uncovered major issues in some cases. The number of excluded cases per dataset and the corresponding exclusion criteria are detailed below:

- NHS: 832 cases were excluded due to severe CMR artefacts affecting the cardiac region in any slice; ICD, pacemaker lead or sternal wires; segmentations including the atria, trabeculations, the LVOT, or the pulmonary valve; or missing or erroneous segmentations in one or more slices.
- Duke: 49 cases were excluded due to erroneous LV and/or RV segmentations. Among those cases, 40 cases were missing a basal segmentation on the LV and/or RV.
- UKBB: 85 cases were excluded due to erroneous LV and/or RV segmentations. Among those cases, 5 cases were missing a basal segmentation on the LV and/or RV.
- ACDC: 18 cases were excluded due to erroneous RV segmentations. Among those cases, 10 cases were missing a basal segmentation on the RV and 8 cases were missing the top basal slice.
- M&Ms: 21 cases were excluded due to erroneous LV segmentations. Among those cases, 15 cases were missing a basal segmentation on the LV and 2 cases were missing the top basal slice.
- M&Ms-2: 15 cases were excluded due to erroneous LV segmentations. Among those cases, 6 cases were missing a basal segmentation on the LV.

For reproducibility purposes, a list of all cases excluded from the online available databases (ACDC, M&Ms and M&Ms2) is below:

| **Database** | **Excluded IDs** |
|---|---|
| ACDC | patient003, patient005, patient006, patient007, patient016, patient023, patient027, patient029, patient035, patient037, patient042, patient047, patient049, patient059, patient074, patient076, patient090, patient094 |
| M&Ms | A4B5U4, A5P5W0, A6B5G9, A7F4G2, C0L7V1, C8O0P2, D1J5P6, D1S5T8, E1L7M3, E7L0N6, E9V9Z2, G9N5V9, H5N0P0, I8Z0Z6, K5M7V5 , L1Q1Z5, N7P3T8, N9P5Z0, O4O6U5, P8V0Y7, Y6Y9Z2 |
| M&Ms-2 | 008, 025, 029, 042, 045, 084, 104, 231, 242, 247, 254, 276, 294, 342, 344 |

## D. Unique CMR input format

Similarly to what is found in the clinic, the datasets used in this study contained different image (DICOM, MHA or NIFTI) and volume (4D or 3D) formats. Therefore, the first pre-processing step was to automatically convert all scans and segmentations to the standard 3D NIFTI format required by the tool. A summary of the diversity of original image and volume formats is shown in Supplemental Table 3.

## E. nnU-net and the adaptive loss function

This automated segmentation framework combines dataset- and expert-driven approaches to decide the optimal framework configuration for a given imaging dataset. The former corresponds to “rule-based” parameters which depend on dataset properties, such as imaging modality or voxel size. The latter corresponds to “fixed parameters” which have been shown to work robustly across a wide range of medical imaging segmentation applications. nnU-Net performs multiple automated pre-processing steps: it crops CMR images to a region of non-zero values, it resamples all voxels to the median voxel spacing, and it performs a z-score normalisation of the intensity values. The median spatial resolution that nnU-Net used was dx, dy, dz = (1.194, 1.194, 8) for X, Y, Z spatial resolution respectively. The same spatial resolution is used at inference time to resample the external databases.

nnU-Net uses the “U-Net” architecture as a template [1]. However, the dataset-specific U-Net configuration - kernel size, number of pooling operations, downsampling/upsampling - is determined by the image size. At this stage, computational resources (i.e., GPU RAM) are also allocated ensuring that the batch size corresponds to less than 5% of voxels in the dataset. The model is trained for 1,000 epochs (where one epoch represents an iteration over 250 mini-batches). Weights are learned via stochastic gradient descent with Nesterov momentum ($\mu = 0.99$) and an initial learning rate of 0.01 with a ‘poly’ learning rate policy $(1 - \text{epoch}/1{,}000)^{0.9}$. Deep supervision (where the contribution towards the total loss after each downsampling is halved) is used, with a combined cross-entropy and Dice loss. To tackle class imbalance, for each training image, 66.7% of patches are from random locations, while 33.3% of patches contain at least one of the foreground classes.

To tackle inconsistent image labelling (e.g., missing LVM on the ES frame) we modified nnU-Net’s loss function. The original loss function calculates the error between the predicted and manual segmentation for each label individually. This means that, when one or more labels are missing from the manual segmentation, even a perfect prediction will result in a non-zero loss. Our modification detects which labels are missing and removes these from the predicted segmentation, thus eliminating their contribution to the overall loss (e.g., if all foreground labels are missing from one patch, the loss is zero and the patch is ignored) [2, 3].

## F. Exclusion of papillary muscles

The LV and RV blood pool segmentations produced by the nnU-Net model are used to mask the image around those regions. Then, Otsu's threshold method [4] is used to exclude the papillary muscles from the LV and RV pool segmentations. The threshold value used for Otsu's method was chosen based on visual inspection of applying this method to the NHS training database. See Supplemental Figure 2 for examples of the original vs post-processed automated segmentations for the Duke dataset.

## G. Post-analysis quality control

Post-analysis quality control of segmentations was based on the following criteria to flag potential errors or unusual cases during automated segmentation:

- one of its ventricular volumes being equal to zero*. These errors were automatically corrected by linear interpolation of the cardiac volume.
- containing outliers (i.e., voxels disconnected from the main segmentation) representing more than 10% of the total number of segmented voxels - this step was first applied to a combination of all segmented labels and then to each individual label*. These errors were automatically corrected by applying largest connected components.
- any label appearing in less than 30% of segmented slices*. These errors were automatically corrected by deleting the corresponding label.
- presence of non-segmented slices between segmented slices for any label*
- negative or zero LV or RV stroke volume*
- LV and RV stroke volume differences larger than 25%*
- LV/RV mass differences between LV and RV larger than 15%
- For full short axis sequences, maximum change of volume between adjacent phases greater than 25 mL

Supplemental Table 4 shows the percentage of automated segmentations that were flagged up by this post-analysis quality control (QC) for each dataset. Overall, cases were mostly flagged up due to large (>25%) stroke volume differences between the LV and the RV.

Note that the criteria marked with asterisks are common between $QA^{gt}$ and QC. We refer the reader to Supplemental Method B for more details of the criteria for automated quality assessment of ground truth segmentations.

## H. Cardiac biomarkers estimation

Cardiac biomarkers were directly computed from segmentation masks for the internal and external database. We use this approach as some biomarkers were not directly available on all external databases. For the NHS training database, we perform a statistical analysis to compare the cardiac biomarkers directly outputted from CVI42 to biomarkers derived from the manual annotated masks and we found no statistically significant difference and a bias less than 5 mL for EDV and ESV and less than 1% for EF.

# Supplemental tables

**Supplemental Table 1 – Details of CMR protocols from different scanners from the NHS database**

| | Siemens Aera 1.5 T | Siemens Biograph mMR 1.5T | Philips Achieva 1.5T/3.0T | Philips Ingenia 1.5T |
|---|---|---|---|---|
| **Scanning sequence** | GR | GR | GR | GR |
| **Sequence variant** | OSP, SK | OSP, SK | SK | SK |
| **Echo time (TE), mm** | 1.03 – 3.17 | 1.44 – 1.50 | 1.28 – 1.67 | 1.27 – 1.55 |
| **Repetition time (TR), mm** | 10.08 – 85.87 | 36.85 – 129.20 | 2.56 – 3.34 | 2.54 – 3.09 |
| **Flip angle, °** | 15 - 57 | 25 | 60 | 10 – 30 |
| **Slice thickness (mm)** | 6 - 10 | 8 | 8 - 10 | 8 - 10 |
| **Pixel spacing, $mm^2$** | 1.52 – 2.39 | 1.41 – 1.56 | 0.87 – 1.46 | 0.86 – 1.34 |
| **Number of frames** | 11 - 50 | 43 - 50 | 15 - 30 | 40 – 50 |

Table shows details of CMR protocols for the different scanner used on the NHS database. Different scanning sequence and sequence variant values are reported in full. Other parameters are as minimum/maximum value.

**Supplemental Table 2 – Training and validation CMR cases**

| Dataset | Train | Validation, before $QA^{gt}$ | Validation, after $QA^{gt}$ |
|---|---|---|---|
| NHS | 2793* | 414* | 414* |
| Duke | 0 | 1319 | 1270 |
| UKBB | 0 | 4872 | 4787 |
| ACDC | 0 | 150 | 132 |
| M&Ms | 0 | 375 | 354 |
| M&Ms-2 | 0 | 360 | 345 |

CMR cases containing segmentations that were used for training and validation (before and after $QA^{gt}$). $QA^{gt}$: data quality assessment of ground truth segmentations.
* see details on the ground-truth data assessment of NHS data in the flow chart in Supplemental Figure 1.

**Supplemental Table 3 – Original CMR image and volume formats for all datasets**

| Dataset | Original image format | | | Original volume format | |
|---|---|---|---|---|---|
| | DICOM | NIFTI | MHA | 4D | 3D |
| NHS | x | | | x | |
| Duke | x | | | x | |
| UKBB | x | | | x | |
| ACDC | | | x | | x |
| M&Ms | | x | | x | |
| M&Ms-2 | | x | | | x |

**Supplemental Table 4 – Automated post-analysis quality control results**

| | Dataset | | | | | |
|---|---|---|---|---|---|---|
| **Post-analysis QC** | **NHS** | **Duke** | **UKBB** | **ACDC** | **M&Ms** | **M&Ms-2** |
| Global outlier >10% [%]* | 0.00 | 0.00 | 0.08 | 0.00 | 0.00 | 0.29 |
| LVBP outlier >10% [%]* | 0.00 | 0.00 | 0.04 | 1.22 | 0.00 | 0.58 |
| LVM outlier >10% [%]* | 0.24 | 0.31 | 0.02 | 0.00 | 0.31 | 0.00 |
| RVBP outlier >10% [%]* | 0.00 | 0.16 | 0.04 | 1.22 | 0.00 | 0.29 |
| SV differences > 25% [%] | 6.35 | 9.51 | 7.08 | 11.22 | 10.99 | 14.78 |
| Flagged up cases [%] | 6.59 | 9.83 | 7.21 | 12.44 | 11.30 | 15.36 |

The first five rows show the percentage of cases that were flagged up for clinician review due to each of the criteria. Since some cases were flagged up after fulfilling multiple criteria, the last row indicates the percentage of cases that were flagged up at least once.

* see Supplemental Method G for a definition of outlier pixels in segmentations.

**Supplemental Table 5 – Inter-observer variability on a subset of the NHS data**

| Absolute errors | O1 vs O2 (n = 50) | O1 vs O3 (n = 50) | O2 vs O3 (n = 50) | Auto vs Manual (n = 50) |
|---|---|---|---|---|
| LVEDV [mL] | 6.98 (7.50) | 6.52 (7.47) | 6.42 (8.47) | 6.98 (7.31) |
| LVESV [mL] | 5.75 (8.29) | 5.68 (9.68) | 7.06 (9.06) | 6.25 (9.01) |
| LVM [g] | 8.78 (8.64) | 8.16 (9.14) | 8.20 (8.96) | 8.88 (8.99) |
| RVEDV [mL] | 8.18 (7.63) | 7.81 (8.25) | 8.31 (9.06) | 8.36 (7.99) |
| RVESV [mL] | 6.10 (8.97) | 5.99 (7.92) | 6.15 (7.79) | 6.5 (8.01) |

Top table: median (interquartile range) absolute errors for cardiac biomarkers for the NHS inter-observer dataset and automated for the same subset of cases used for NHS inter-observer variability. LVBP: left ventricular blood pool, MYO: left ventricular myocardium, RVBP: right ventricular blood pool, LVBP: LV blood pool, MYO: LV myocardium, RVBP: RV blood pool, ED: end diastole, ES: end systole.

**Supplemental Table 6 – Dice scores per scanner model**

| | LVBP | | MYO | | RVBP | |
|---|---|---|---|---|---|---|
| **Scanner model** | **ED** | **ES** | **ED** | **ES** | **ED** | **ES** |
| Philips Achieva (1.5T, 3.0T) (n=297) | 0.95 (0.02) | 0.91 (0.05) | 0.85 (0.05) | 0.85 (0.05) | 0.93 (0.03) | 0.88 (0.05) |
| Siemens Aera (1.5T) (n=4936) | 0.94 (0.02) | 0.88 (0.05) | 0.82 (0.05) | 0.84 (0.04) | 0.90 (0.03) | 0.83 (0.06) |
| Siemens Avanto (1.5T) (n=272) | 0.93 (0.04) | 0.87 (0.07) | 0.82 (0.04) | 0.83 (0.05) | 0.91 (0.04) | 0.85 (0.08) |
| Siemens Biograph mMR (3.0T) (n=7) | 0.95 (0.01) | 0.91 (0.03) | 0.86 (0.03) | N/A* | 0.92 (0.02) | 0.87 (0.04) |
| General Electric Excite (1.5T) (n=58) | 0.95 (0.02) | 0.89 (0.07) | 0.85 (0.05) | 0.87 (0.04) | 0.92 (0.04) | 0.88 (0.06) |
| General Electric HDxt (3.0T) (n=14) | 0.95 (0.02) | 0.92 (0.04) | 0.83 (0.03) | 0.86 (0.03) | 0.89 (0.04) | 0.86 (0.05) |
| Philips Ingenia (1.5T) (n=114) | 0.95 (0.02) | 0.92 (0.04) | 0.86 (0.04) | 0.84 (0.05) | 0.92 (0.03) | 0.88 (0.05) |
| Canon Orian (1.5T) (n=25) | 0.94 (0.03) | 0.91 (0.04) | 0.83 (0.03) | 0.86 (0.03) | 0.92 (0.03) | 0.91 (0.03) |
| Siemens Sola (1.5T) (n=146) | 0.93 (0.03) | 0.86 (0.08) | 0.82 (0.04) | 0.86 (0.04) | 0.91 (0.03) | 0.86 (0.05) |
| Siemens Symphony (1.5T) (n=101) | 0.95 (0.02) | 0.92 (0.05) | 0.86 (0.05) | 0.86 (0.05) | 0.91 (0.06) | 0.87 (0.06) |
| Siemens Trio (3.0T) (n=3) | 0.97 (0.00) | 0.92 (0.04) | 0.87 (0.01) | 0.86 (0.02) | 0.89 (0.03) | 0.82 (0.08) |
| Siemens Verio (3.0T) (n=47) | 0.94 (0.02) | 0.84 (0.09) | 0.82 (0.03) | N/A* | 0.91 (0.04) | 0.86 (0.05) |
| Siemens Vida (3.0T) (n=581) | 0.93 (0.02) | 0.86 (0.07) | 0.83 (0.03) | 0.85 (0.03) | 0.91 (0.04) | 0.86 (0.05) |

Median (interquartile range) values for each scanner model and label (LVBP, MYO, and RVBP) in ED and ES. *No manual segmentations contained the myocardium in ES. LVBP: LV blood pool, MYO: LV myocardium, RVBP: RV blood pool, ED: end diastole, ES: end systole.

**Supplemental Table 7 – Comparison of our proposed model with and without QA, and against original nnUNet model trained only on the M&Ms data.**

| Model | Dataset | LVBP [%] | MYO [%] | RVBP [%] | Average [%] |
|---|---|---|---|---|---|
| Ours with $QA^{gt}$ | NHS | 94.3 (4.0) | 85.5 (4.4) | 90.8 (5.4) | 91.3 (7.5) |
| Ours without $QA^{gt}$ | NHS | 93.1 (4.76)* | 83.6 (5.25)* | 89.2 (4.43)* | 88.6 (7.9)* |
| M&Ms | NHS | 92.2 (5.1)* | 81.7 (4.7)* | 88.1 (5.7)* | 87.3 (8.2)* |

Dice scores are shown as median (interquartile range) percentages for each the NHS validation dataset, including values per label and their average. First row shows the results for our proposed method with $QA^{gt}$, second row shows the results for our proposed method without $QA^{gt}$ and third row shows the results for the comparative state-of-the-art method trained only on the M&M database. Comparisons between the Dice scores of the NHS validation cases for our proposed method with $QA^{gt}$ and the two comparative approaches (without $QA^{gt}$ and original M&M model) were performed using Mann-Whitney U tests. Asterisks indicate statistically significant differences for each label, where * = $p < 0.01$. LVBP: LV blood pool, MYO: LV myocardium, RVBP: RV blood pool.

**Supplemental Table 8 – Validation experiments: Dice scores per dataset with and without QC.**

| | Dice scores without QC | | | | Dice scores with QC | | | |
|---|---|---|---|---|---|---|---|---|
| **Dataset** | **LVBP** | **MYO** | **RVBP** | **Average** | **LVBP** | **MYO** | **RVBP** | **Average** |
| NHS | 93.49 (3.76) | 84.66 (4.25) | 89.80 (4.43) | 90.22 (5.27) | 94.3 (3.01) | 85.5 (4.22) | 90.8 (4.41) | 91.3 (5.24) |
| Duke | 89.22 (6.80)* | 82.20 (4.09)* | 87.90 (5.92)* | 87.72 (6.52)* | 91.3 (6.50) | 83.0 (3.81) | 89.3 (5.83) | 89.2 (6.33) |
| UKBB | 90.79 (4.81) | 82.53 (4.86) | 86.64 (5.89) | 86.65 (6.21) | 91.8 (4.92) | 83.0 (4.53) | 87.8 (5.82) | 87.4 (5.95) |
| ACDC | 93.40 (5.38)* | 86.39 (4.69)* | 89.60 (6.63)* | 89.80 (6.31) | 95.5 (4.91) | 87.4 (3.52) | 91.8 (6.04) | 90.6 (5.95) |
| M&Ms | 91.62 (5.72)* | 84.51 (4.82)* | 88.38 (6.90) | 88.17 (6.56) | 93.4 (5.41) | 85.4 (4.71) | 90.4 (5.83) | 89.4 (6.34) |
| M&Ms-2 | 93.51 (4.02)* | 85.03 (5.17) | 89.76 (5.47) | 89.44 (6.02) | 94.6 (4.01) | 86.0 (5.02) | 90.9 (5.73) | 90.3 (6.31) |

Left table: median (interquartile range) values for each validation dataset, including values per label and their average, excluding cases that did not pass the $QA^{gt}$. Right table: median (interquartile range) values for each validation dataset, including values per label and their average, including cases that passed the $QA^{gt}$. LVBP: left ventricular blood pool, MYO: left ventricular myocardium, RVBP: right ventricular blood pool, $QA^{gt}$: data quality assessment of ground truth segmentations. Comparisons between the Dice scores with and without $QA^{gt}$ were performed using Mann-Whitney U tests. Bonferroni correction for multiple comparisons. Asterisks indicate statistically significant differences for each label after correction (20 tests), where * = $p < 0.01/20$.

**Supplemental Table 9 – Absolute errors for cardiac biomarkers derived from the automated and manual segmentations with and without QC**

| | Absolute errors for cardiac biomarkers without QC | | | | | | | Absolute errors for cardiac biomarkers with QC | | | | | | |
|---|---|---|---|---|---|---|---|---|---|---|---|---|---|---|
| | LV | | | | RV | | | LV | | | | RV | | |
| **Dataset** | **EDV [mL]** | **ESV [mL]** | **EF [%]** | **LVM [g]** | **EDV [mL]** | **ESV [mL]** | **EF [%]** | **EDV [mL]** | **ESV [mL]** | **EF [%]** | **LVM [g]** | **EDV [mL]** | **ESV [mL]** | **EF [%]** |
| NHS | 8.96 (8.04) | 7.97 (6.93) | 3.38 (2.83) | 11.83 (9.93) | 11.06 (9.43) | 8.49 (7.41) | 4.17 (3.39) | 6.7 (13.5) | 6.3 (12.9) | 3.4 (4.3) | 8.9 (17.9) | 8.5 (15.8) | 6.4 (12.5) | 4.2 (5.4) |
| Duke | 7.25 (6.62) | 6.62 (6.24) | 5.32 (4.73) | 16.75 (13.70)* | 8.49 (8.58) | 6.33 (6.17)* | 5.37 (4.55) | 5.7 (11.0) | 5.0 (9.7) | 5.3 (7.1) | 13.3 (17.2) | 6.0 (11.8) | 4.5 (9.1) | 5.2 (6.8) |
| UKBB | 9.36 (7.98)* | 9.39 (6.62)* | 4.67 (3.39) | 7.82 (6.27)* | 11.51 (9.37)* | 10.64 (8.42)* | 5.31 (3.92) | 7.3 (12.5) | 8.4 (9.6) | 4.7 (4.2) | 6.6 (12.8) | 9.2 (14.7) | 8.9 (12.5) | 5.3 (5.9) |
| ACDC | 4.51 (3.39) | 7.51 (6.32)* | 4.56 (4.31) | 9.04 (8.25)* | 8.17 (7.54) | 9.94 (10.44)* | 6.00 (5.28) | 3.6 (5.8) | 5.3 (8.0) | 4.6 (5.1) | 6.8 (12.7) | 6.1 (11.2) | 5.7 (11.1) | 5.6 (7.6) |
| M&Ms | 9.18 (8.78)* | 7.00 (6.29)* | 4.49 (3.93) | 10.73 (10.42)* | 11.21 (10.24) | 8.82 (10.83)* | 5.66 (5.22) | 6.2 (13.4) | 5.1 (9.8) | 4.2 (5.4) | 8.2 (16.6) | 8.4 (16.9) | 5.9 (13.7) | 5.6 (7.3) |
| M&Ms-2 | 8.29 (7.46)* | 7.22 (6.29) | 3.47 (3.18) | 9.39 (8.53)* | 13.00 (13.98) | 11.68 (12.20)* | 5.98 (4.98) | 6.3 (9.6) | 5.5 (8.4) | 3.4 (4.2) | 7.3 (14.2) | 8.7 (16.5) | 8.3 (13.9) | 5.9 (7.0) |
| Average | 7.93 | 7.62 | 4.32 | 10.93 | 10.57 | 9.32 | 5.42 | 5.97 | 5.93 | 4.27 | 8.52 | 7.82 | 6.62 | 5.30 |

Left: Median (interquartile range) absolute errors for cardiac biomarkers derived from the automated and manual segmentations for each validation dataset, excluding cases that did not pass the $QA^{gt}$. Right: Median (interquartile range) absolute errors for cardiac biomarkers derived from the automated and manual segmentations for each validation dataset, including cases that passed the $QA^{gt}$. $QA^{gt}$: data quality assessment of ground truth segmentations. Last row shows a summary of mean error values across all datasets. Comparisons between absolute errors for cardiac biomarkers with and without $QA^{gt}$ were performed using Mann-Whitney U tests. Bonferroni correction for multiple comparisons. Asterisks indicate statistically significant differences for each label after correction (35 tests), where * = $p < 0.01/35$.

# Supplemental figures

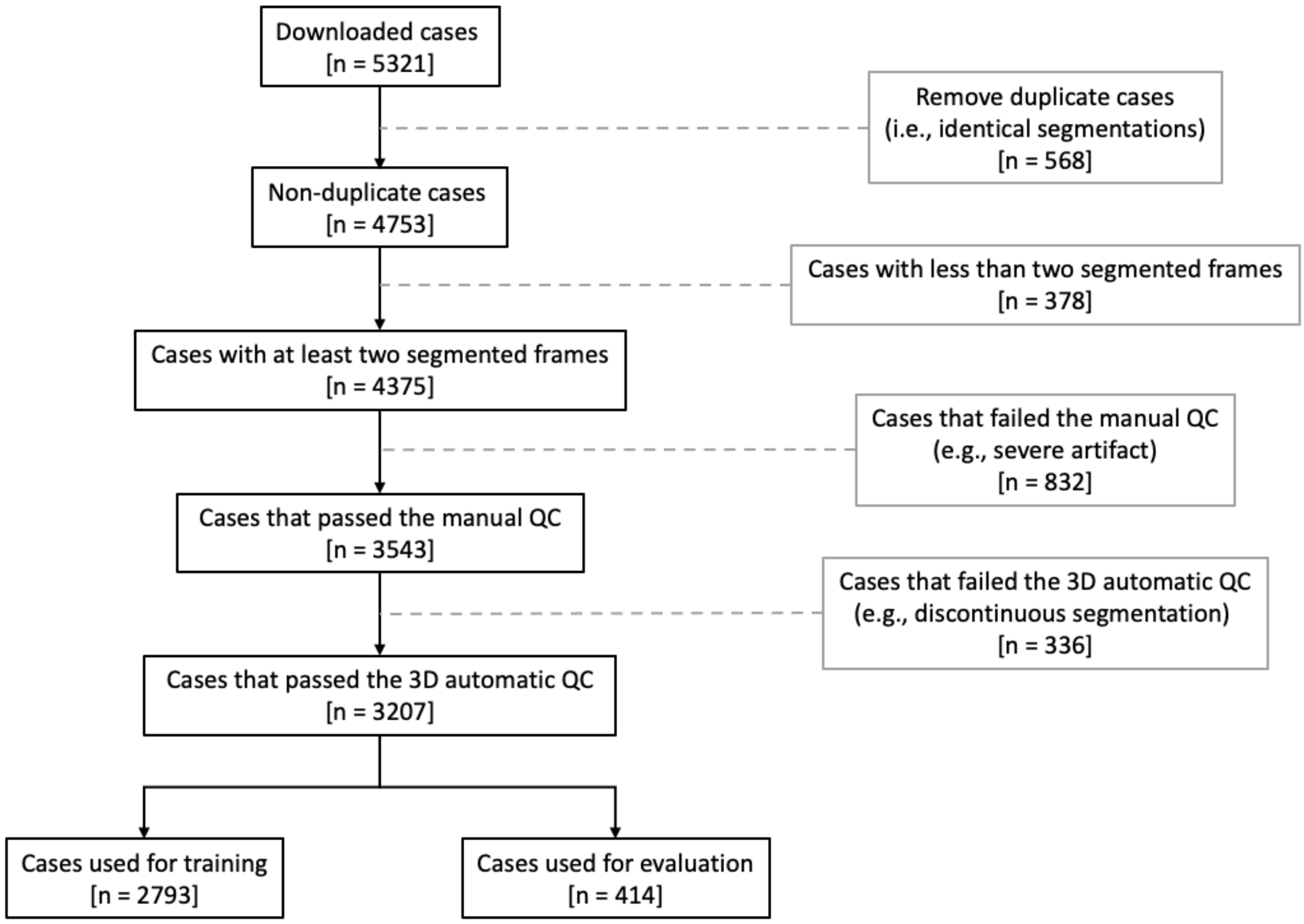

**Supplemental Figure 1 – Internal (NHS) dataset curation process**: Number of NHS cases discarded during download, pre-processing, and quality assessment ($QA^{gt}$).

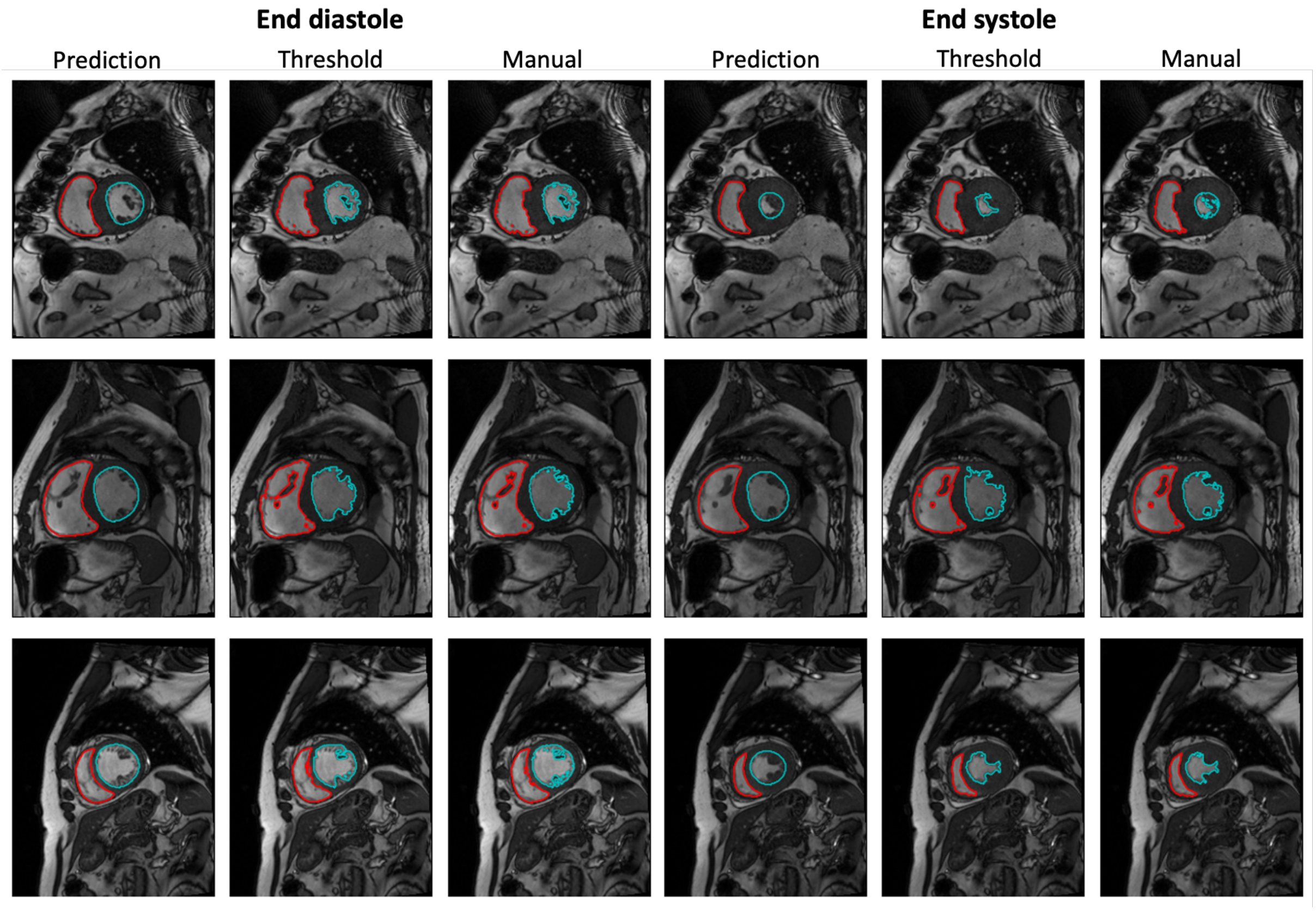

**Supplemental Figure 2 – Original vs post-processed automated segmentations for the Duke dataset**: Comparison between original and post-processed automated segmentations for the Duke dataset using Otsu's threshold method. Each row depicts end-diastolic (left) and end-systolic (right) segmentations located in the middle of the heart for a random case.

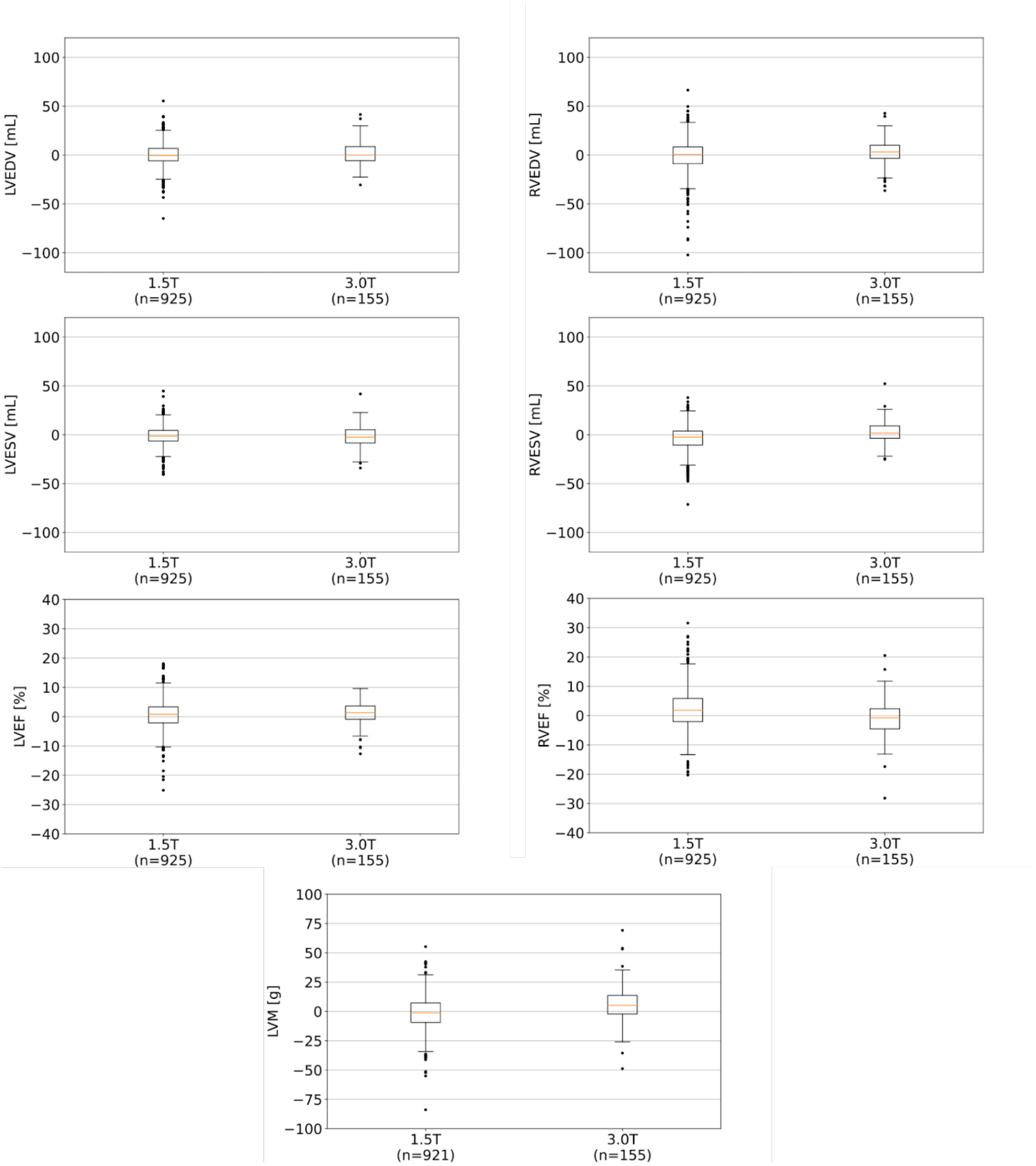

**Supplemental Figure 3 – Box plots of cardiac biomarkers per field strength group**: Box plots of manually and automatically derived cardiac biomarkers for each magnetic field strength group. Abbreviations as in Figure 2.

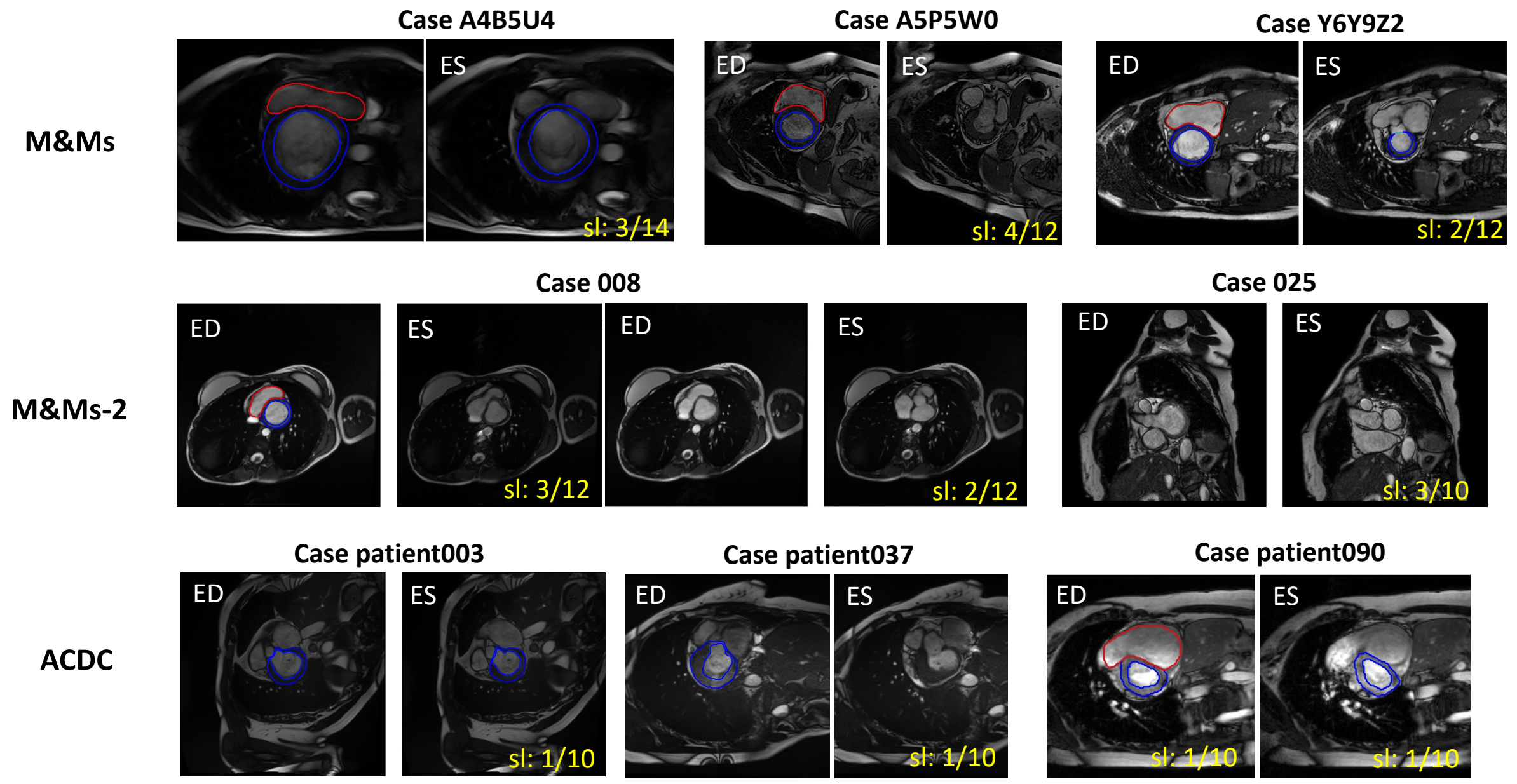

**Supplemental Figure 4 – Example of ground-truth cases from external databases flagged by our quality assessment of ground-truth data ($QA^{gt}$).** Selection of excluded cases for the online external databases (M&Ms, M&Ms-2 and ACDC). The majority of cases were excluded because of missing basal slices, with the short-axis stack not continuing into the atria (examples show the top slices of case A5P5W0 from M&Ms and case patient090 from ACDC). Other issues raised were missing segmentation (see the clear myocardial rim for both LV and RV in the ES in slice 2 and ED in slice 3 of case 008 from M&Ms-2, missing RV segmentation in the ES frame from Patient090, LV segmentation in the ES frame of case A5P5W0, the missing RVOT segmentation in the ES frame of case Y6Y972 and the LV segmentation in the ED frame of case 025 from M&Ms-2), lastly we automatically raised SV differences of larger than 25% between RV and LV (here case A4B5UA, from M&Ms)

# References

1. Ronneberger O, Fischer P, Brox T. U-net: Convolutional networks for biomedical image segmentation. In: Lecture Notes in Computer Science (including subseries Lecture Notes in Artificial Intelligence and Lecture Notes in Bioinformatics). 2015. p. 234–41.
2. Tilborghs S, Bertels J, Robben D, Vandermeulen D, Maes F. The Dice Loss in the Context of Missing or Empty Labels: Introducing Phi and epsilon. 2022. p. 527–37.
3. Petit O, Thome N, Charnoz A, Hostettler A, Soler L. Handling Missing Annotations for Semantic Segmentation with Deep ConvNets. Deep Learning in Medical Image Analysis and Multimodal Learning for Clinical Decision Support: 4th International Workshop, DLMIA 2018, and 8th International Workshop, ML-CDS 2018, Held in Conjunction with MICCAI 2018, Granada, Spain, September 20, 2018, P; 2018. p. 20–8.
4. Otsu N. A Threshold Selection Method from Gray-Level Histograms. IEEE Trans Syst Man Cybern. 1979;C:62–6.